\documentclass[a4paper]{jpconf}
\usepackage{graphicx}
\usepackage{iopams}
\usepackage{amsmath}
\usepackage{cite}
\usepackage[pdftex]{hyperref}

\newcommand{\avg}[1]{\langle #1 \rangle}

\begin{document}

\title{Rapidity correlations test stochastic hydrodynamics}

\author{C Zin$^1$, S Gavin$^1$ and G Moschelli$^2$}

\address{$^1$ Department of Physics and Astronomy, Wayne State University, 
Detroit, MI, 48202\\
$^2$ Lawrence Technological University, 21000 West Ten Mile Road, Southfield, MI  48075}

\ead{christopherzin@wayne.edu, sean.gavin@wayne.edu, gmoschell@ltu.edu}


\begin{abstract}
We show that measurements of the rapidity dependence of transverse momentum correlations can be used to determine the characteristic time $\tau_{\pi}$ that dictates the rate of isotropization of the stress energy tensor, as well as the shear viscosity $\nu = \eta/sT$. We formulate methods for computing these correlations using second order dissipative hydrodynamics with noise. Current data are consistent with $\tau_{\pi}/\nu \sim 10$ but targeted measurements can improve this precision.
\end{abstract}


\section{Introduction}
\label{sec_intro}
From experiments at the Relativistic Heavy Ion Collider and the Large Hadron Collider we have seen that hydrodynamic models can be used to describe the expansions of a range of collision systems. In large nucleus-nucleus systems these descriptions were used as evidence of the production of a liquid-like quark gluon plasma. Smaller proton-proton and proton-nucleus collisions lack the temperature and density necessary to create quark gluon plasma so experiment was met with surprise when hydrodynamic models were successfully applied to these systems. To investigate the differences between these systems we turn our interest to the processes by which these systems thermalize.

Our focus is on the rapidity dependence of $p_t$ correlations which give information about shear viscosity and the transport coefficient $\tau_\pi$, the relaxation time for pressure isotropization. We approach the problem by using a stochastic description of hydrodynamics to develop differential equations for the momentum correlation function of the system. In doing so, we highlight the importance of including stochastic processes in the description of an evolving collision system. Our results, which are compared with experiment, show that momentum correlations are propagated through the system via second order diffusion. For a full discussion of the material in these proceedings, see \cite{Gavin:2016hmv}.


\section{Brownian Motion}
\label{sec_brown}
In heavy ion collisions, the initial phase space distribution $f = dN/d^3p \, d^3x$ fluctuates from event to event due to impact parameter variations and the distributions of the colliding nuclei. Stochastic noise introduces important two-particle correlations to the system which we model after the familiar example of Brownian motion\footnote{A detailed discussion of Brownian motion and stochastic processes can be found in \cite{gardiner2004handbook}.}. In Brownian motion, a heavy particle suspended in a fluid obeys the Langevin equation
\begin{equation} \label{eqn_langevin}
\Delta v = -\gamma v(t) \Delta t + \Delta W
\end{equation}
where $v(t)$ is the velocity of the heavy particle and $\Delta v = v(t + \Delta t) - v(t)$ is the change in velocity during the time step $\Delta t$. The first term on the right hand side of Eq. (\ref{eqn_langevin}) represents the viscous drag force felt by the heavy particle, having strength $\gamma$. The second term is the contribution to $\Delta v$ from stochastic noise which represents collisions of the heavy particle with the fluid medium. 

One assumes that collisions are random in both magnitude and direction and that the variance of the noise is proportional to $\Delta t$. We represent this as $\avg{\Delta W} = 0$ and $\avg{\Delta W^2} = \Gamma \Delta t$, where $\Gamma$ determines the strength of the noise. Notice that as a result of this assumption $\Delta W$ becomes an infinitesimal of order $\Delta t/2$.

Taking the average of Eq. (\ref{eqn_langevin}) and the limit $\Delta t \rightarrow 0$ gives a differential equation for $\avg{v}$
\begin{equation} \label{eqn_avg-langevin}
\Delta \avg{v} = -\gamma \avg{v(t)} \Delta t + \avg{\Delta W} \quad \Rightarrow \quad d\avg{v}/dt = -\gamma \avg{v}
\end{equation}
which is free of noise. The effect of the noise is felt by $\avg{v^2(t)}$, however, as we can see by averaging the square of Eq. (\ref{eqn_langevin}) and taking $\Delta t \rightarrow 0$: 
\begin{equation} \label{eqn_avg-v-sq}
\Delta \avg{v^2} = 2\avg{v \Delta v} + \avg{\Delta v \Delta v} = -2\gamma \avg{v^2} \Delta t + \Gamma \Delta t \quad \Rightarrow \quad d\avg{v^2}/dt = -2\gamma \avg{v^2} + \Gamma.
\end{equation}
The first equality in Eq. (\ref{eqn_avg-v-sq}) is known as the It$\hat{\mathrm o}$ product rule in stochastic calculus and is required so as to keep all terms of order $\Delta t$.

Our goal is to obtain a deterministic equation for the square of the mean velocity. To this end, notice that in equilibrium the time derivative in Eq. (\ref{eqn_avg-v-sq}) must vanish, which can be seen by applying the equipartition theorem to $\avg{v^2}_{eq}$. This leaves $\Gamma = 2 \gamma \avg{v^2}_{\mathrm eq} = 2 \gamma T/m$ where the second equality follows again through use of the equipartition theorem. Determining the strength of the noise in this manner is an application of the fluctuation-dissipation theorem. In the context of Brownian motion, the fluctuation-dissipation theorem says the energy of the Brownian particle which is dissipated via friction is added to the medium. In turn, the fluctuations in the medium increase, which is felt by the Brownian particle via more energetic collisions.

Notice, at this point, the importance of including stochastic noise in our calculation. Had we ignored the noise term in Eq. (\ref{eqn_avg-v-sq}) we would have found that $\avg{v^2(t)}$ vanishes in equilibrium, a clear violation of the equipartition theorem. By including stochastic noise, through use of the Langevin equation, we have allowed our results to be consistent with well established physics.

Returning to Eq. (\ref{eqn_avg-v-sq}), we define the variance of the velocity as $r = \avg{v^2} - \avg{v}^2$ and the deviation of $r$ from its equilibrium value as $\Delta r = r - r_{eq}$. We then find $\Delta r$ satisfies $d \Delta r/dt = -2 \gamma \Delta r$. This is a completely deterministic equation for $\Delta r$ as the noise was absorbed into $r_{eq}$ by application of the fluctuation-dissipation theorem. Also we emphasize that $\Delta r$ relaxes with timescale $1/2\gamma$, half the value of the relaxation of $\avg{v}$ as seen in Eq. (\ref{eqn_avg-langevin}). These points are important for our main results later on.


\section{Hydrodynamic Fluctuations}
\label{sec_hydro}
We turn now to heavy ion collisions. After a collision fluctuations in density and temperature cause the transverse velocity of neighboring system cells to vary along the beam axis. Viscosity works to diffuse fluctuations in the velocity throughout the system. This spread in fluctuations is dominated by shear hydrodynamic modes. Small fluctuations in velocity produce correspondingly small fluctuations in the momentum current $\mathbf{M} = w \mathbf{v}$ where $w = e + p$ is the enthalpy density with energy density $e$ and pressure $p$. This momentum density satisfies the linear first order Navier-Stokes equation
\begin{equation} \label{eqn_navier-stokes}
\frac{\partial}{\partial t} \mathbf{M} + \nabla p = \frac{\zeta + \frac{1}{3} \eta}{w} \nabla ( \nabla \cdot \mathbf{M} ) + \frac{\eta}{w} \nabla^2 \mathbf{M}
\end{equation}
where $\zeta$ and $\eta$ are the bulk and shear viscosities. To separate out the shear modes we perform a Helmholtz decomposition on $\mathbf{M}$ by writing $\mathbf{M} = \mathbf{g_l} + \mathbf{g}$ where the longitudinal modes satisfy $\nabla \times \mathbf{g_l} = 0$ and the shear modes satisfy $\nabla \cdot \mathbf{g} = 0$. Using this, we find the shear modes obey the diffusion equation
$\partial \mathbf{g}/\partial t = \nu \nabla^2 \mathbf{g}$ for the kinematic viscosity $\nu = \eta/w$.

We apply our treatment of Brownian motion to the shear modes by writing this as a difference equation with stochastic noise: $\Delta g_i = \nu \nabla^2 g_i \Delta t + \Delta W_i$. Here, the noise satisfies $\avg{\Delta W_i} = 0$ and $\avg{\Delta W_i(x_1) \Delta W_j(x_2)} = \Gamma^{ij}_{12} \Delta t$. We then define the correlation function $r = \avg{g_i(x_1) g_j(x_2)} - \avg{g_i(x_1)} \avg{g_j(x_2)}$ and the deviation of $r$ from its equilibrium value $\Delta r = r - r_{eq}$ where $r_{eq} = wT \delta_{ij} \delta(x_1 - x_2)$ is found through application of the fluctuation-dissipation theorem. This allows us to write down an explicit form for the noise $\Gamma^{ij}_{12} = -(\nabla^2_1 + \nabla^2_2) r_{eq}$.
Notice that due to the $\delta$-functions in $r_{eq}$ there is no noise-noise correlation between the different components of $\mathbf{g}$ nor at different spatial positions $x_1 \neq x_2$.

We can now write an evolution equation for the momentum correlations of the system
\begin{equation} \label{eqn_1st-order-diff-deltar}
\left[ \frac{\partial}{\partial \tau} - \frac{\nu}{\tau^2} \left( \frac{\partial^2}{\partial \eta^2_1} + \frac{\partial^2}{\partial \eta^2_2} \right) \right] \Delta r = 0
\end{equation}
where $\tau$ is the proper time and $\eta$ is the spatial rapidity. Thus, correlations of the shear modes spread through the system via first order diffusion. Unfortunately, first order diffusion is acausal so a $\delta$-function spike in momentum will become a Gaussian, instantaneously spreading its tails to infinity. To remedy this issue we must repeat this procedure in second order hydrodynamics.

Starting with a second order M$\ddot{\mathrm u}$ller-Israel-Stewart equation we find that the shear modes satisfy a Maxwell-Cattaneo equation to linear order
\begin{equation} \label{eqn_max-cat}
\left[ \tau_\pi \frac{\partial^2}{\partial t^2} + \frac{\partial}{\partial t} \right] \mathbf{g} = \nu \nabla^2 \mathbf{g}.
\end{equation}
This equation introduces the transport coefficient $\tau_\pi$, the relaxation time for isotropization of the stress-energy tensor. Applying our Brownian prescription to Eq. (\ref{eqn_max-cat}), we find a deterministic evolution equation for the correlations of the shear modes
\begin{equation} \label{eqn_main-result}
\left[ \frac{\tau^*_\pi}{2} \frac{\partial^2}{\partial \tau^2} + \frac{\partial}{\partial \tau} - \frac{\nu^*}{\tau^2} \left( \frac{\partial^2}{\partial \eta^2_1} + \frac{\partial^2}{\partial \eta^2_2} \right) \right] \Delta r = 0
\end{equation}
where the, now time dependent, coefficients are given by $\tau^*_\pi = \tau_\pi / (1 + \kappa \tau_\pi/\tau)$, $\nu^* = \nu / (1 + \kappa \tau_\pi/\tau)$ and $\kappa = (1 + d \ln (\tau_\pi/\eta T)/d \ln \tau)/2$. For simplicity we take these to be constant in this work.

Equation (\ref{eqn_main-result}) has two important features we wish to emphasize. First, the combination of wave-like and diffusion-like behavior given by the first two terms renders the theory causal, correcting the problem of Eq. (\ref{eqn_1st-order-diff-deltar}). Second, due to the factor of $1/2$ in the first term, $\Delta r$ relaxes at half the rate of the average, as seen in Eq. (\ref{eqn_max-cat}). This is the same type of behavior we saw in the case of Brownian motion.



\begin{figure}[t]
\centering
\includegraphics[scale=0.8]{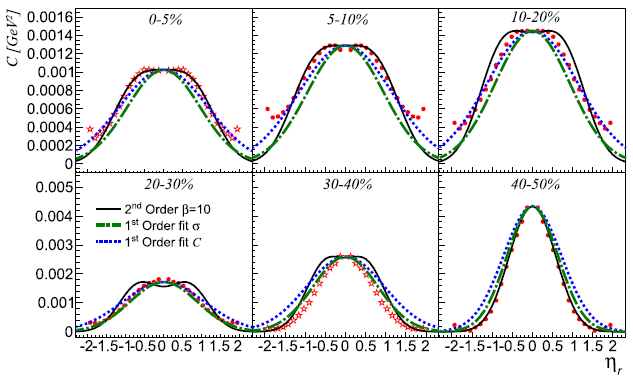}
\caption{First and second order calculations of momentum correlations in relative rapidity $\eta_r = \eta_2 - \eta_1$. First order lines are fit to the rapidity width $\sigma$ of $\Delta r$ and directly to the shape of $\mathcal{C}$. Data are from \cite{Agakishiev:2011fs} (open stars) and \cite{PrivComm} (filled circles). Percentages indicate centrality bins.}
\label{fig_C}
\end{figure}

For comparison to experiment we choose the transverse momentum covariance
\begin{equation}
\mathcal{C} = \avg{N}^{-2} \left\langle \textstyle{\sum_{j \neq i} p_{ti}} \, p_{tj} \right\rangle - \avg{p_t}^2
\end{equation}
where $\avg{N}$ is the average multiplicity and $\avg{p_t} = \avg{\sum_i p_{ti}}/\avg{N}$ is the average transverse momentum. This observable has been used to study viscosity \cite{Agakishiev:2011fs} and has a particularly simple connection to $\Delta r$ given by $\avg{N}^2 \, \mathcal{C} = \int \Delta r \, dx_1 dx_2$. We study $\mathcal{C}$ in both the first and second order theory by using Eq. (\ref{eqn_1st-order-diff-deltar}) and Eq. (\ref{eqn_main-result}), with constant coefficients, to calculate $\Delta r$. Our results are shown in Fig. (\ref{fig_C}).

The experimental data has a Gaussian profile in the peripheral centrality bins. In more central bins the data develop a double bump structure for points close in relative rapidity. Our first order lines fit the Gaussian profile of the peripheral bins but maintain their Gaussian shape as we move to more central bins, losing agreement with the data. We see that as the second order line evolves to more central collisions the peak flattens to hit the broad shoulders developed by the data. Overall we find that the second order line better agrees with experiment.


\section{Conclusions}
\label{sec_conc}
Our goal was to develop equations for the evolution of the correlations of the shear momentum modes in a heavy ion collision system. We made an effort to study correlation functions relative to their values in equilibrium in order to account for the stochastic nature of particle collisions. Our main result, Eq. (\ref{eqn_main-result}), describes the evolution of momentum correlations in second order hydrodynamics. We find that experimental data cannot be adequately explained using first order hydrodynamics. Second order hydrodynamics better explains the features shown by experiment. This introduces the relaxation time $\tau_\pi$, which sets the rate for pressure isotropization of the system. Using our best fits to experiment we find $\tau_\pi/\nu \sim 10$, double the value predicted by kinetic theory. We feel this estimate can be improved by using more realistic parameters in Eq. (\ref{eqn_main-result}) but save this for future work.


\ack
We thank Rajendra Pokharel for contributing to the early stages of this project as well as Monika Sharma and Claude Pruneau for discussing the STAR data. This work was supported in part by the U.S. NSF grant PHY-1207687. 


\section*{References}

\bibliographystyle{iopart-num}
\bibliography{zin_hq2016_bib}

\end{document}